\documentclass[conference]{IEEEtran}
\IEEEoverridecommandlockouts

\usepackage{cite}
\usepackage{amsmath,amssymb,amsfonts}
\usepackage{algorithmic}
\usepackage{graphicx}
\usepackage{textcomp}
\usepackage{xcolor}
\usepackage{float}
\usepackage{booktabs}
\usepackage{url}
\usepackage{newunicodechar}
\newunicodechar{≈}{\approx}
\usepackage{stfloats}

\def\BibTeX{{\rm B\kern-.05em{\sc i\kern-.025em b}\kern-.08em
    T\kern-.1667em\lower.7ex\hbox{E}\kern-.125emX}}

\title{LLM-Based Code Documentation Generation and Multi-Judge Evaluation}

\author{
    \IEEEauthorblockN{Ikbel Ghrab\textsuperscript{1,2}, Mohamed Dhieb\textsuperscript{2}, Ismail Khenissi\textsuperscript{2}, Ines Abdeljaoued-Tej\textsuperscript{1,3}}\\
    \IEEEauthorblockA{\textsuperscript{1}University of Carthage, Engineering School of Statistics and Information Analysis (ESSAI), Ariana, Tunisia}\\
    \IEEEauthorblockA{\textsuperscript{2}Easy Transfer, 72 Rue de la Division Leclerc, 91160 Saulx-les-Chartreux, Paris, France}\\
    \IEEEauthorblockA{\textsuperscript{3}University of Tunis El Manar, Laboratory of Bioinformatics, bioMathematics, and bioStatistics (LR16IPT09), \\
    Institut Pasteur de Tunis, 13 Place Pasteur, B.P. 74, Belvédère, 1002 Tunis, Tunisia}
    
}

\begin{document}

\IEEEpubid{%
  \makebox[\columnwidth]{%
    \parbox{\columnwidth}{%
      \scriptsize
      979-8-3315-2212-4/25/\$31.00~\copyright2026 IEEE. Personal use of this material is permitted. Permission from IEEE must be obtained for all other uses, in any current or future media, including reprinting/republishing this material for advertising or promotional purposes, creating new collective works, for resale or redistribution to servers or lists, or reuse of any copyrighted component of this work in other works.
    }%
  }%
  \hspace{\columnsep}%
  \makebox[\columnwidth]{}
}
\maketitle
\IEEEpubidadjcol

\begin{abstract}
High-quality source code documentation is vital yet often neglected, especially in critical domains like healthcare where reliability and maintainability are essential. We presented an AI-powered framework that automates documentation generation from code and repositories using eight state-of-the-art Large Language Models (LLMs), including GPT, Gemini, Qwen, and LLaMA variants. Built on the PocketFlow orchestration framework, the system applies modular pipelines and advanced prompt engineering to produce structured, context-aware documentation. To ensure quality and guide model selection, we introduced a Multi-LLM-as-Judges evaluation framework, where four independent LLMs assess outputs across nine criteria, such as Completeness, Clarity, and Faithfulness. Experiments conducted on an open-source medical physics library—demonstrated showed a 42\% performance gap between top and bottom models. By combining diverse model outputs, optimized prompting, and rigorous evaluation, our approach enhances documentation quality and reduces manual effort—especially in safety-critical healthcare software.

\end{abstract}

\section{Introduction}

High-quality software documentation is essential in healthcare, where reliability, transparency, and compliance directly impact patient safety. Medical software—ranging from imaging tools to treatment planning systems—requires clear documentation to ensure developers, clinicians, and auditors can understand underlying code logic, data flow, and decision mechanisms \cite{sanders2008dealing,van2011complementing,tan2024detecting}. However, creating and maintaining such documentation manually is time-consuming, error-prone, and difficult to scale as systems evolve \cite{manshreck2020documentation,zhao2024evaluating,dagenais2010creating}.  
Recent advances in Large Language Models (LLMs) offer new opportunities to automate and enhance documentation workflows \cite{jelodar2025llms}. Yet, existing tools—such as DocuWriter.ai \cite{docuwriterai}, Mintlify \cite{mintlify}, and Workik \cite{workik}—often generate superficial summaries, overlook architectural complexity, and lack standardized evaluation methods \cite{chakrabarty2024customizable}.  Moreover, selecting the most suitable LLM for domain-specific documentation, especially in regulated fields like healthcare, therefore remains an open challenge requiring systematic benchmarking and context-aware assessment.

This paper presents an AI-powered documentation generation system designed to address these limitations. The system employs a modular pipeline architecture based on the PocketFlow orchestration framework \cite{zhang2025flowstatehumansenabling}, incorporating advanced prompt engineering techniques to ensure technical precision and narrative clarity. To overcome model selection uncertainty, we integrate a novel Multi-LLM-as-Judges evaluation paradigm, wherein documentation outputs generated by eight LLMs—spanning the GPT, Gemini, Qwen, and LLaMA families — are independently assessed by four additional LLMs acting as evaluators. This evaluation framework scores each output across nine dimensions: Completeness, Clarity, Coherence, Faithfulness, Hallucination Detection, Code Quality, Mermaid Quality, Relevance, and Accuracy.

\section{Pipeline Solution Architecture}


The automated documentation generation system employs a modular pipeline architecture built upon the Pocketflow framework, designed to transform raw source code repositories into structured, pedagogically sound tutorial documentation. The system addresses the fundamental challenge of bridging the gap between complex implementation details and accessible learning materials by implementing a progressive refinement model where each processing stage builds upon the outputs of its predecessors to create increasingly sophisticated representations of the codebase's conceptual structure.


The pipeline architecture promotes maintainability, testability, and flexibility through its decomposition of the complex documentation generation task into discrete, manageable processing nodes. Each node performs a specific function within the overall workflow, processing data and passing results to subsequent stages while maintaining state consistency across the entire pipeline. This modular design enables easy modification or extension of individual processing steps without disrupting the overall system architecture, while the integration of Large Language Models throughout the pipeline enables sophisticated tasks such as source code understanding, abstraction identification, semantic relationship extraction, and natural language generation of tutorial content.

A foundational component of this system is the deliberate and systematic use of prompt engineering to guide and constrain LLM behavior within each pipeline node. Rather than issuing generic queries, the pipeline employs specialized prompt templates aligned with the specific functional and semantic role of each node. These prompts are designed to emulate professional personas—such as software architects, systems modelers, technical writers, and instructional designers—while also encoding strict output format requirements and semantic expectations. This architectural commitment to prompt engineering ensures both the interpretability and reliability of LLM-generated artifacts throughout the pipeline.

\subsection{Data Acquisition and Preprocessing Pipeline}

The initial stage of the pipeline centers on the \texttt{FetchRepo} node, which is responsible for retrieving, filtering, and standardizing source code files from both local directories and remote GitHub repositories. It abstracts the differences between source types, enforcing a uniform data representation across the pipeline. While this node does not engage in direct LLM interaction, it plays a critical role in preparing contextual anchors that inform downstream prompt engineering. By generating deterministic file indices, maintaining relative path consistency, and extracting lightweight metadata, it enables subsequent nodes to construct LLM prompts that reference code in a compact, positionally stable manner, thereby reducing ambiguity and enhancing interpretability.

\subsection{Semantic Analysis and Abstraction Identification}

The \texttt{IdentifyAbstractions} node marks the entry point of language model interaction. It utilizes LLMs to extract conceptual abstractions from the codebase, such as modules, classes, data flows, and major system features. The prompt engineering strategy for this stage instructs the language model to assume the role of a senior software architect tasked with distilling a software system’s conceptual scaffolding for onboarding or documentation purposes. The prompt presents the model with a structured preamble, indexed code summaries, and explicit instructions to generate YAML-formatted abstraction entries, each with a human-friendly name, description, and associated file indices. These prompts are carefully tuned to balance coverage with conciseness, and to elicit abstractions that reflect design intent rather than surface-level syntax. Multiple iterations were conducted to ensure that the LLM reliably avoids redundancy, respects schema constraints, and aligns abstractions with real architectural patterns observed in professional software systems.

\subsection{Relationship Analysis and Architectural Modeling}

Following abstraction identification, the \texttt{AnalyzeRelationships} node deepens the system’s semantic understanding by modeling how abstractions interact. The prompt here is constructed to simulate the analytical reasoning of a system integration expert. It combines previously generated abstraction metadata with annotated code excerpts, and requests the language model to infer and describe architectural relationships between components. The model is instructed to detect usage patterns, data dependencies, method invocations, or inheritance structures, and to express each connection using structured descriptors that identify the source, target, and nature of the relationship. Prompt engineering at this stage focuses heavily on clarity and disambiguation, especially when interpreting code fragments that may suggest multiple types of relationships. Special attention is given to prompt phrasing that avoids spurious links and encourages semantically grounded analysis. Each invocation ensures that the entire abstraction graph is connected, producing a complete and interpretable architectural model.

\subsection{Pedagogical Structuring and Content Organization}

The \texttt{OrderChapters} node transforms the architectural model into an instructional sequence by determining the most pedagogically effective order in which to present the identified abstractions. To accomplish this, the associated prompt casts the language model as an educational designer developing a progressive curriculum for learners encountering the system for the first time. The prompt includes the full list of abstractions, their interdependencies, and descriptive metadata, and instructs the model to propose a logical order that minimizes cognitive load and supports concept-building. Emphasis is placed on introducing foundational topics before dependent or advanced ones, while maintaining thematic continuity across the sequence. The language model is constrained to output a linear, non-redundant ordering that respects all prerequisite relationships. The prompt's internal guidance also discourages grouping abstractions arbitrarily, ensuring that the resulting chapter structure supports coherent, narrative-driven documentation.

\subsection{Content Generation and Tutorial Synthesis}

The core content of the tutorial is produced by the \texttt{WriteChapters} node, which uses LLMs to generate full-length explanatory chapters for each abstraction. The prompt at this stage is the most complex and heavily engineered in the pipeline. It presents the language model with a rich contextual package that includes the abstraction definition, related architectural relationships, project-level summaries, and annotated code examples. The LLM is instructed to adopt the role of a technical writer with a mandate to create beginner-accessible, professionally structured instructional content. The prompt includes explicit expectations for output structure, such as introductory context, motivational framing, detailed explanation, and conclusions. It also requests embedded code snippets and Mermaid diagrams to illustrate structural and behavioral aspects of the abstraction. This level of prompt detail is essential to achieving high-quality content that balances technical rigor with readability. Repeated testing and refinement of prompt variants were necessary to ensure consistent chapter formatting, tone alignment, and conceptual progression throughout the documentation set.

\subsection{Integration and Final Assembly}

The \texttt{CombineTutorial} node finalizes the documentation by assembling all chapter outputs into a cohesive, navigable resource. This node performs structural integration rather than language generation, but it relies on the uniformity and consistency enforced by previous prompt engineering decisions. Because all earlier nodes use well-specified output formats and narrative conventions, this stage is able to construct cross-references, table of contents entries, and chapter links automatically. Optional LLM prompts may be used here to generate global introductory sections or summaries, again using minimal, targeted prompt templates that frame the documentation at the project level without duplicating existing content. Overall, the success of this integration stage depends on the upstream prompt engineering discipline that ensures predictable, composable outputs.

\subsection{Prompt Engineering as an Architectural Layer}

Throughout the pipeline, prompt engineering plays a central, architectural role—on par with traditional software concerns such as modularity, state management, and data flow. Each node’s functionality is inseparable from the design of its associated prompt, which defines the LLM’s persona, objective, input framing, output schema, and stylistic tone. These prompts are not one-off ad hoc strings, but engineered artifacts refined through empirical iteration, internal tooling support, and user-in-the-loop validation. By encoding intent and structure directly into the language interface, the system transforms prompt design into a first-class architectural concern—enabling robust, interpretable, and high-fidelity automation of complex documentation tasks.

\section{Innovative Scoring and Aggregation Methodology}
\label{sec:enhanced-scoring-methodology}

Selecting an appropriate large language model (LLM) for automated code documentation requires both rigorous evaluation and cross-model consistency. To address this, a scalable benchmarking framework was developed and applied to a real-world healthcare software repository—\textit{PyMedPhys}, an open-source medical physics library.\footnote{\url{https://github.com/pymedphys/pymedphys}} This project’s multidisciplinary nature, combining clinical computation, data management, and medical device interfacing, provides a robust and domain-relevant foundation for evaluating LLM documentation capabilities.

Eight LLMs across four major model families (\texttt{Gemini}, \texttt{GPT}, \texttt{Qwen}, and \texttt{LLaMA}) were employed to generate and evaluate documentation. A \textit{multi-LLM-as-a-Judge} ensemble strategy was used, wherein outputs produced by each model were independently reviewed by four judge models from the same families. This ensemble approach minimizes model-specific biases, enhances robustness, and enables consensus-based quality estimation across diverse architectures.

Documentation quality was evaluated using nine weighted metrics: \textbf{Accuracy}, \textbf{Faithfulness}, \textbf{Completeness}, \textbf{Clarity}, \textbf{Coherence}, \textbf{Relevance}, \textbf{Code Example Quality}, \textbf{Mermaid Diagram Quality}, and \textbf{Hallucination} \cite{laskar2023large,khan2024evaluation,malin2025review}.  
Hallucination carried a heavier penalty due to its critical impact on factual reliability, especially in healthcare-related contexts. The aggregation process incorporated three key methodological innovations: (1) a hierarchical penalty mechanism emphasizing hallucination suppression, (2) a content density analysis to reward structured and contextually balanced outputs, and (3) a weighted scoring scheme prioritizing core metrics such as Accuracy and Faithfulness.  
Final scores were computed through consensus-based fusion across all judge evaluations, with significant inter-model disagreements automatically flagged for further analysis.

\subsection{Detailed Scoring Pipeline}

The scoring methodology processes evaluations through a sophisticated multi-stage pipeline designed to capture both quantitative metrics and qualitative assessment nuances:

\paragraph{Stage 1: Consensus Score Calculation}
For each documentation chapter, individual scores for each metric from the four judge models are aggregated to establish a consensus baseline:
\begin{eqnarray}
\text{\bf RawScore}_m &=& \frac{1}{4} \sum_{j=1}^{4} \text{\bf Score}_{j,m}
\end{eqnarray}

where $\text{\bf Score}_{j,m}$ represents the score assigned by judge $j$ for metric $m$, and $\text{\bf RawScore}_m$ is the resulting consensus score for that metric.

\paragraph{Stage 2: Score Normalization}
Raw scores are transformed to a standardized 0-1 scale to enable consistent weighted aggregation:
\begin{eqnarray}
\text{\bf NormalizedScore}_m &=& \frac{\text{\bf RawScore}_m - 1}{4}
\end{eqnarray}

This normalization ensures that all metrics contribute proportionally to the final evaluation, regardless of their original scale characteristics.

\paragraph{Stage 3: Hallucination Penalty Application}
Given the critical importance of factual accuracy in technical documentation, the hallucination metric receives specialized treatment through a two-step penalty mechanism:
\begin{eqnarray}
\text{\bf InvertedScore}_h &=& 1 - \text{\bf NormalizedScore}_h \\
\text{\bf PenalizedScore}_h &=& \text{\bf InvertedScore}_h \times \lambda_h
\end{eqnarray}

where $\lambda_h = 2.0$ represents the hallucination penalty multiplier, effectively doubling the impact of detected fabricated content on the final score.

\paragraph{Stage 4: Weighted Metric Integration}
The processed scores are combined using a carefully calibrated weighting scheme that reflects the relative importance of different quality dimensions:
\begin{eqnarray}
\text{\bf WeightedScore} &=& \sum_{m \in \mathcal{M}} \text{\bf ProcessedScore}_m \times w_m
\end{eqnarray}

where $\mathcal{M}$ represents the set of all evaluation metrics, $\text{\bf ProcessedScore}_m$ is either the normalized score or penalized score (for hallucination), and $w_m$ denotes the weight assigned to metric $m$. The weight distribution prioritizes core technical qualities while maintaining attention to auxiliary elements:
\begin{itemize}
    \item \textbf{Core Metrics} (Accuracy, Faithfulness, Completeness): $w_{core} = 0.55$ (combined)
    \item \textbf{Secondary Metrics} (Clarity, Coherence, Relevance): $w_{secondary} = 0.30$ (combined)
    \item \textbf{Auxiliary Metrics} (Code Quality, Mermaid Quality): $w_{auxiliary} = 0.10$ (combined)
    \item \textbf{Hallucination Penalty}: $w_{hallucination} = 0.05$ (with amplification)
\end{itemize}

\paragraph{Stage 5: Content Density Analysis}
The framework incorporates an automated assessment of documentation structure through content density analysis:
\begin{equation}
\text{\bf Density}_{avg} = \frac{\sum_{s=1}^{S} \text{WordCount}_s}{S}
\end{equation}

where $S$ represents the total number of sections in the documentation and $\text{WordCount}_s$ is the word count for section $s$. Based on this analysis, a density weight $w_{density}$ is assigned according to the following thresholds:
\begin{align*}
w_{density} = \begin{cases}
0.7 & \text{if } \text{\bf Density}_{avg} < 50 \text{ (very sparse)} \\
0.8 & \text{if } 50 \leq \text{\bf Density}_{avg} < 100 \text{ (sparse)} \\
1.2 & \text{if } 100 \leq \text{\bf Density}_{avg} < 200 \text{ (balanced)} \\
1.0 & \text{if } 200 \leq \text{\bf Density}_{avg} < 300 \text{ (dense)} \\
0.9 & \text{if } \text{\bf Density}_{avg} \geq 300 \text{ (very dense)}
\end{cases}
\end{align*}

\paragraph{Stage 6: Final Score Computation}
The weighted score is adjusted by the content density factor and converted back to the original 1-5 scale:
\begin{eqnarray}
\text{\bf FinalScore}_{0-1} &=& \text{\bf WeightedScore} \times w_{density} \\
\text{\bf FinalScore}_{1-5} &=& 1 + (\text{\bf FinalScore}_{0-1} \times 4)
\end{eqnarray}

\subsection{Documentation Set Aggregation}

Individual chapter scores are aggregated at the documentation set level to provide comprehensive quality assessments:

\begin{equation}
\text{\bf SetScore}_{avg} = \frac{1}{N} \sum_{c=1}^{N} \text{\bf FinalScore}_{1-5,c}
\end{equation}

where $N$ represents the number of chapters in the documentation set and $\text{\bf FinalScore}_{1-5,c}$ is the final score for chapter $c$. This scoring methodology introduces several key innovations that significantly enhance the reliability and validity of AI-generated documentation evaluation. \textbf{Multi-Dimensional Quality Assessment}: Unlike binary or single-metric approaches, the framework captures multiple facets of documentation quality, providing a comprehensive view of model performance across diverse evaluation criteria. \textbf{Adaptive Penalty Mechanisms}: The hallucination penalty system creates a non-linear response to factual errors, ensuring that even minor inaccuracies receive appropriate attention while severely penalizing systematic fabrication. \textbf{Structural Quality Integration}: The content density analysis provides automated assessment of documentation organization, rewarding well-structured content without requiring manual evaluation of organizational quality. \textbf{Judge Consensus Robustness}: The multi-judge approach mitigates individual model biases and provides more stable, reproducible evaluations compared to single-judge systems. \textbf{Scalable Evaluation Pipeline}: The framework's modular design enables efficient evaluation of large documentation sets while maintaining consistent quality assessment standards. 

\section{Results}
\label{sec:results}

\begin{table*}[!t]
\caption{Overall Scores for Documentation Generation Models (Weighted Evaluation)}
\begin{center}
\begin{tabular}{|c|c|c|c|c|c|c|c|c|}
\hline
\textbf{Metric} & \textbf{\shortstack{gemini\\2.5-pro}} & \textbf{\shortstack{gemini\\2.5-flash}} & \textbf{\shortstack{gpt\\4.1-nano}} & \textbf{\shortstack{gpt\\5-nano}} & \textbf{\shortstack{llama\\3.2-11b}} & \textbf{\shortstack{llama\\4-scout}} & \textbf{\shortstack{qwen3\\32b}} & \textbf{\shortstack{qwen3\\30b-a3b}} \\
\hline
\textbf{Avg. FinalScore} & \textbf{4.68} & \textbf{4.65} & 3.15 & 3.70 & 2.85 & 2.70 & 3.36 & 3.55 \\
\hline
\textbf{Avg. RawScore} & 4.52 & 4.45 & 3.46 & 3.89 & 3.15 & 2.90 & 3.44 & 3.50 \\
\hline
Completeness & 4.40 & 4.38 & 3.70 & 3.97 & 3.27 & 2.80 & 3.58 & 3.67 \\
\hline
Clarity & 4.80 & 4.36 & 3.97 & 4.50 & 3.92 & 3.55 & 4.03 & 4.17 \\
\hline
Code Quality & 4.55 & 4.47 & 3.28 & 3.82 & 2.85 & 2.57 & 3.30 & 3.40 \\
\hline
Mermaid Quality & 4.67 & 4.65 & 3.67 & 4.36 & 2.12 & 1.67 & 3.10 & 2.17 \\
\hline
Coherence & 4.87 & 4.95 & 4.07 & 4.73 & 4.15 & 3.80 & 4.25 & 4.32 \\
\hline
Faithfulness & 4.40 & 4.45 & 3.58 & 3.16 & 2.75 & 2.50 & 3.15 & 3.37 \\
\hline
Hallucination & 3.52 & 3.30 & 1.23 & 2.43 & 2.22 & 2.80 & 2.20 & 2.67 \\
\hline
Accuracy & 4.63 & 4.55 & 3.71 & 3.55 & 3.05 & 3.02 & 3.20 & 3.42 \\
\hline
Relevance & 4.88 & 4.97 & 3.90 & 4.53 & 4.02 & 3.42 & 4.15 & 4.32 \\
\hline
\multicolumn{9}{l}{$^{\mathrm{a}}$Final scores integrate weighted metrics and hallucination penalties.}
\end{tabular}
\label{tab:overall_scores_summary}
\end{center}
\end{table*}

The evaluation results reveal clear performance tiers among the tested models (see Table~\ref{tab:overall_scores_summary}).  
The average \textbf{final scores}, computed under the weighted metric system, distinctly differentiate model quality and behavior.  
\textbf{Top Tier:} \texttt{gemini-2.5-pro} (4.68) and \texttt{gemini-2.5-flash} (4.65), both demonstrating strong completeness, structural clarity, and low hallucination rates.  
\textbf{Middle Tier:} \texttt{gpt-5-nano} (3.70) and \texttt{qwen3-30b-a3b} (3.55), which maintain solid coherence and relevance despite some verbosity or reduced faithfulness.  
\textbf{Lower Tier:} \texttt{gpt-4.1-nano} (3.15), \texttt{qwen3-32b} (3.36), and both \texttt{llama} models (2.70–2.85), which lag behind primarily due to weaker code quality, factual instability, and higher hallucination impact.  
The 1.98-point spread between the top and lowest models reflects substantial qualitative variation across the evaluated systems.

\subsection{Weighted vs. Raw Score Analysis}
\label{subsec:weighted-raw-analysis}

Comparing weighted and raw averages provides insights into the role of metric weighting and hallucination penalties in the ranking outcomes.  
For \textbf{Gemini models}, weighted scores slightly exceed raw averages, suggesting robust balance and minimal penalization across heavily weighted metrics such as Accuracy, Faithfulness, and Completeness.  
In contrast, \textbf{GPT models} display more complex behavior: while \texttt{gpt-5-nano} maintains stability, \texttt{gpt-4.1-nano} suffers a sharp decline from its raw score (3.46) to the weighted average (3.15), primarily due to hallucination incidence (1.23) and over-dense content distribution.  
The \textbf{LLaMA models} consistently drop below their raw averages, indicating compounding penalties from both factual inaccuracies and poor mermaid/code quality.  
\textbf{Qwen models} remain balanced, with minimal differences between weighted and raw scores, confirming their “normal” output profile—adequate but not exceptional in either direction.

\subsubsection{Core Metrics (Combined Weight: 55\%)}
\label{subsubsec:core-metrics}

Core metrics (Accuracy, Faithfulness, Completeness) heavily influenced final rankings.  
\texttt{gemini-2.5-pro} achieved consistently high marks (4.63 Accuracy, 4.40 Faithfulness, 4.40 Completeness), followed closely by \texttt{gemini-2.5-flash} with a similarly stable profile.  
\texttt{gpt-5-nano} performed respectably (3.55 Accuracy, 3.16 Faithfulness, 3.97 Completeness) but exhibited uneven factual grounding.  
Both \texttt{llama} models scored below 3.0 across faithfulness and completeness, demonstrating weak grounding and fragmented document structure.  
\texttt{qwen3-30b-a3b} maintained moderate reliability (3.4–3.7 range), consistent with its middle-tier classification.

\subsubsection{Secondary Metrics (Combined Weight: 30\%)}
\label{subsubsec:secondary-metrics}

Secondary metrics (Clarity, Coherence, Relevance) emphasize the structural and semantic quality of documentation.  
Both Gemini variants again led, with \texttt{gemini-2.5-flash} achieving near-perfect coherence (4.95) and relevance (4.97).  
The \texttt{gpt-5-nano} model produced coherent and logically structured outputs (4.73 Coherence) but at times suffered from information overload—consistent with its “content-dense” generation style.  
Qwen models remained clear and relevant, while the \texttt{llama} family displayed inconsistent clarity and lower relevance scores, reflecting less controlled output structure.

\subsubsection{Auxiliary Metrics (Combined Weight: 10\%)}
\label{subsubsec:auxiliary-metrics}
Auxiliary dimensions capture technical and structural fidelity, notably code and Mermaid diagram quality.  
The Gemini models again dominated with near-equal code ($\approx 4.5$) and Mermaid ($\approx 4.6$) performance.

\texttt{gpt-5-nano} followed closely (4.36 Mermaid), indicating good diagrammatic alignment.  
The \texttt{llama} models demonstrated clear limitations, particularly in diagram rendering (1.67–2.12).  
Qwen models achieved reasonable but unremarkable technical precision.

\subsubsection{Hallucination Assessment (Weight: 5\%)}
\label{subsubsec:hallucination-assessment}

Hallucination scores provide a crucial measure of factual stability.  
Gemini maintained moderate-to-low hallucination rates (3.52 and 3.30), while GPT variants varied substantially: \texttt{gpt-4.1-nano} suffered from significant factual drift (1.23), contrasting with the more stable \texttt{gpt-5-nano} (2.43).  
LLaMA and Qwen families clustered around moderate values (2.2–2.8), reflecting general reliability but occasional factual inconsistencies.  
These penalties explain part of the gap between raw and weighted outcomes for models with otherwise acceptable raw averages.

\subsection{Judge Consensus and Consistency}
\label{subsec:judge-consensus}

The collective evaluation confirms the reliability of the multi-LLM judging framework, with high inter-model agreement and limited scoring variance across evaluators.  
Models demonstrated distinct behavioral patterns reflecting trade-offs between density, precision, and factual stability.  
\texttt{gemini-2.5-pro} and \texttt{gemini-2.5-flash} consistently sustained balanced, factually grounded outputs across all weighted dimensions, explaining their strong stability under penalized scoring.  
\texttt{gpt-5-nano} and \texttt{gpt-4.1-nano} showed greater dispersion—producing detailed yet sometimes overextended documentation that incurred penalties under faithfulness and hallucination weighting.  
Both \texttt{llama-3.2-11b} and \texttt{llama-4-scout} displayed limited informational depth and weaker structural fidelity, while the \texttt{qwen3} series maintained moderate, evenly distributed results.  

Across evaluators, score distributions revealed minimal disagreement on model ordering, confirming robustness of the ensemble-based assessment.  
Overall, models achieving equilibrium between semantic richness and factual precision—most notably the Gemini variants—exhibited not only higher mean scores but also the lowest inter-metric variance, underscoring the role of balanced generation strategies in producing consistent documentation performance.

\section{Discussion and Perspectives}
\label{subsec:limitations}

This study was limited to a single project within the healthcare domain and relied exclusively on freely accessible LLM APIs. These constraints restricted experimentation with high-performance commercial models and prevented large-scale evaluations across diverse programming languages, frameworks, and project complexities. The evaluation also lacks comprehensive validation on other domain-specific repositories, such as healthcare informatics platforms or safety-critical systems, which may require more specialized documentation strategies and domain-informed model behavior.  

Moreover, the current evaluation framework depends solely on automated LLM-based assessment without the inclusion of human expert review. Integrating domain specialists—such as clinicians for healthcare software or regulatory experts for compliance-critical systems—could provide richer contextual insight and improve the interpretive reliability of results. The absence of such expert input remains a key limitation, particularly for assessing documentation quality in specialized or high-stakes fields.  

The models tested in this study were general-purpose LLMs without domain-specific fine-tuning. This likely limited their ability to produce documentation that is both contextually accurate and compliant with domain regulations. In healthcare-related settings, for instance, achieving high-quality documentation requires not only linguistic precision but also a solid understanding of medical terminology, regulatory requirements, and clinical workflow logic.  
Despite these limitations, the proposed multi-LLM evaluation framework represents a foundational contribution toward systematic, cross-domain benchmarking of documentation generation quality. Future research directions include the following:

\begin{itemize}
    \item \textbf{Broader Validation:} Expanding evaluations to encompass diverse programming languages, frameworks, and large-scale open-source systems (e.g., Apache, Linux kernel) to improve generalizability and cross-domain robustness.
    
    \item \textbf{Domain Fine-Tuning:} Adapting and fine-tuning LLMs for specialized domains—particularly healthcare—to enhance factual precision, regulatory compliance, and contextual awareness. Incorporating domain-specific corpora and standards (e.g., FDA 21 CFR Part 820, HL7 FHIR) can enable generation of compliant, domain-aware documentation.
    
    \item \textbf{Expert-Informed Evaluation:} Integrating domain experts into the assessment process to validate LLM-based evaluations, refine scoring methodologies, and curate high-quality training data for future judge model development.
    
    \item \textbf{Healthcare Specialization:} Prioritizing documentation tasks related to clinical workflows, safety reviews, and regulatory audits. The development of healthcare-specific evaluation models trained on domain literature could further improve assessment precision.
    
    \item \textbf{Framework Expansion:} Extending the current evaluation framework with domain-sensitive metrics, user feedback integration, and CI/CD compatibility to enable continuous, automated monitoring of documentation quality.
\end{itemize}


\section{Conclusion}
\label{sec:conclusion}

This work tackled the manual burden of source code documentation by developing a framework that leverages a novel \textit{multi-LLM-as-Judges} paradigm in which multiple large language models act simultaneously as generators and evaluators. The framework was applied to \textit{PyMedPhys}, an open-source medical physics library, using nine weighted metrics to evaluate documentation completeness, clarity, factual accuracy, and coherence. Results revealed distinct performance tiers: \texttt{gemini-2.5-pro} (4.68) and \texttt{gemini-2.5-flash} (4.65) consistently achieved the highest weighted scores, followed by \texttt{gpt-5-nano} (3.70) and \texttt{qwen3-30b-a3b} (3.55), while \texttt{llama} variants performed comparatively lower due to weaker structural and factual consistency. The strong inter-model agreement and low scoring variance confirm the robustness of the proposed multi-judge approach. Overall, this study establishes a solid foundation for scalable, cross-model benchmarking of documentation generation, with promising potential for domain-specific adaptation in healthcare and other regulated fields.

\section*{Acknowledgment}

We would like to thank the anonymous reviewers for their insightful feedback, constructive suggestions, and encouraging remarks, all of which significantly contributed to enhancing the quality and clarity of this work.

\bibliographystyle{IEEEtran}
\bibliography{references}

\end{document}